\documentclass[a4paper,12pt]{article}
\usepackage[utf8]{inputenc}
\usepackage{mathtools} 
\usepackage{amssymb, amsmath, amsfonts, amsthm, bm}

\usepackage{natbib}
\def\spacingset#1{\renewcommand{\baselinestretch}%
{#1}\small\normalsize} \spacingset{1}
\spacingset{1.65}

\usepackage{hyperref}

\usepackage{graphicx} 
\graphicspath{{Figures/}} 
\usepackage{subcaption} 

\newcommand{\mbbR}{\mathbb{R}}

\newcommand{\mbfY}{\mathbf{Y}}
\newcommand{\bfmu}{\bm{\mu}}
\newcommand{\bfbeta}{\bm{\beta}}
\newcommand{\mcalN}{\mathcal{N}}
\newcommand{\matern}{{Mat\'{e}rn}}

\newcommand{\bmY}{\bm{Y}}

\newcommand{\HSigma}{\scalebox{1.5}{$\Sigma$}}

\newcommand{\bms}{\bm{s}}

\newcommand{\mbbE}{\mathbb{E}}

\newcommand{\mcalD}{\mathcal{D}}

\usepackage{authblk}

\title{Asymmetric Cross-Correlation in Multivariate Spatial Stochastic Processes: A Primer}
\author{Xiaoqing Chen}
\affil[]{Department of Mathematics and Statistics, 
University of Exeter, Exeter, EX4 4PY,  U.K.}
\affil[]{Email: xiaoqing.a.chen@gmail.com}
\author{}
\date{}

\begin{document}

\maketitle

\begin{abstract}
Multivariate spatial phenomena are ubiquitous, spanning domains such as climate, pandemics, air quality, and social economy. 
Cross-correlation between different quantities of interest at different locations is asymmetric in general. 
This paper provides the visualization, structure, and properties of asymmetric cross-correlation as well as symmetric auto-correlation.
It reviews mainstream multivariate spatial models and analyzes their capability to accommodate asymmetric cross-correlation. It also illustrates the difference in model accuracy with and without asymmetric accommodation using a 1D simulated example.
\end{abstract}

\section{Introduction}
\label{sec:intro}
Multivariate spatial phenomena are ubiquitous, spanning diverse disciplines. Examples include climate change (e.g., temperature, precipitation, wind speed),  global pandemic (e.g., protein mutation rate, UV radiation intensity, regional vaccination coverage), air quality (e.g., PM2.5, NO$_2$, O$_3$), and social economy (e.g., crime rates, housing prices, income levels), to name a few.

The common feature of these spatial phenomena is that each quantity of interest at a certain location $s_i$ not only interacts with itself at nearby locations in the domain $\mathcal{D}$ but also interacts with other quantities at both the same location $s_i$ and nearby locations in $\mathcal{D}$. 

For example, the concentration level of PM2.5 at a given location $s_i$, 
not only interacts with PM2.5 at a nearby location $s_j$, but also interacts with other pollutants, such as NO$_2$, O$_3$, at the same location $s_i$ as well as at a
nearby location $s_j$ in $\mathcal{D}$.


There are three types of interactions here. The first type involves 
the same component at $s_i$ and its nearby location $s_j$, which is captured by \textit{same-component auto-correlation} (or \textit{same-component auto-covariance}).
The second type involves two different components at the same location $s_i$, and is reflected by 
\textit{same-location auto-correlation} (or \textit{same-location auto-covariance}). 
The third type involves two different components at $s_i$ and its nearby location $s_j$, and is captured by \textit{cross-correlation} (or \textit{cross-covariance}).

Auto-correlations, whether same-component or same location, are all \textit{symmetric}. 
Using PM2.5 as an example, the correlation between PM2.5 at $s_i$ and PM2.5 at $s_j$ 
is the same as the correlation between PM2.5 at $s_j$ and PM2.5 at $s_i$. 
Similarly, the correlation between PM2.5 and NO$_3$ at $s_i$ is the same as the correlation between NO$_3$ and PM2.5 at $s_i$.
These symmetries can be easily verified using mathematical expressions: 
$cov(PM2.5(s_i), PM2.5(s_j)) \equiv cov(PM2.5(s_j), PM2.5(s_i))$ and  
$cov(PM2.5(s_i), O_3(s_i)) \equiv cov(O_3(s_i), PM2.5(s_i))$, as per the definition of covariance.

In contrast, cross-correlation is \textit{asymmetric} in general. 
For example, consider PM2.5 and Sea Salt (SS). 
The cross-correlation between PM2.5 at one location (e.g., Saudi Arabia) and SS at another location (e.g., Egypt) is not the same as the cross-correlation between PM2.5 in Egypt and  SS in Saudi Arabia. 
Mathematically, $cov(X(s_i), Y(s_j)) \neq cov(X(s_j), Y(s_i))$, or $cov(X(s_i), Y(s_j)) \neq cov(Y(s_i), X(s_j))$,
where $X$ can represent PM2.5 and $Y$ can represent SS.

\begin{figure}[ht]
  \centering
  \begin{subfigure}[b]{0.244\textwidth}
    \caption{Lon:$[-180^{\circ}, -90^{\circ})$}
    \includegraphics[width=0.95\textwidth]{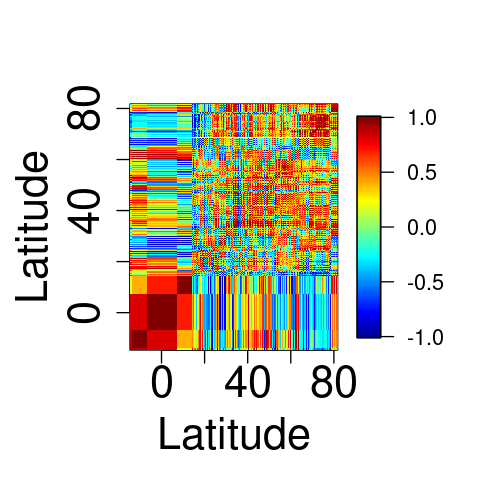}
  \end{subfigure}
  \begin{subfigure}[b]{0.24\textwidth}
    \caption{Lon:$[-90^{\circ}, 0^{\circ})$}
    \includegraphics[width=0.95\textwidth]{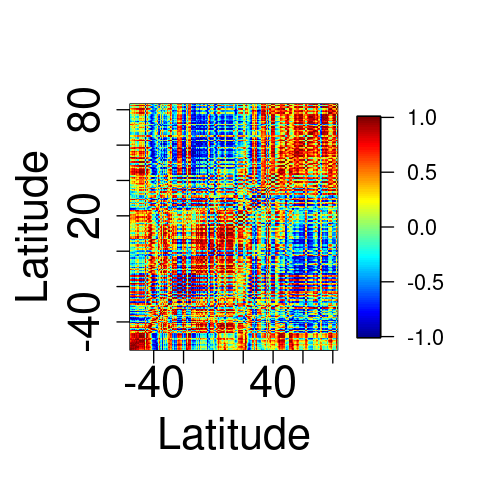}
  \end{subfigure}
  \begin{subfigure}[b]{0.24\textwidth}
    \caption{Lon:$[0^{\circ}, 90^{\circ})$}
    \includegraphics[width=0.95\textwidth]{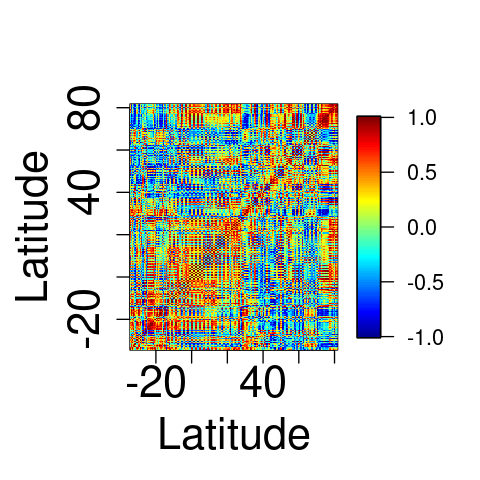}
  \end{subfigure}
  \begin{subfigure}[b]{0.24\textwidth}
    \caption{Lon:$[90^{\circ}, 180^{\circ}]$}
    \includegraphics[width=0.95\textwidth]{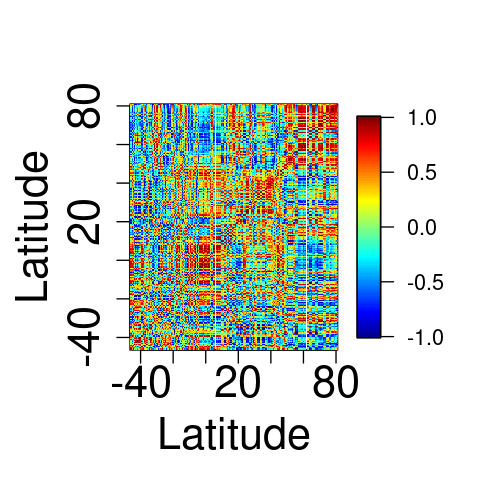}
  \end{subfigure}
  \caption{Empirical same-component auto-correlation matrix plots for PM2.5 across four longitude strips $[-180^{\circ}, -90^{\circ})$, $[-90^{\circ}, 0^{\circ})$, $[0^{\circ}, 90^{\circ})$, $[90^{\circ}, 180^{\circ}]$. All symmetric about y = x. }
  \label{fig:auto_corr_PM25}
\end{figure}

Figure \ref{fig:auto_corr_PM25} shows the empirical same-component auto-correlation matrix plot of PM2.5, displaying $corr(PM2.5(s_i), PM2.5(s_j)) \equiv corr(PM2.5(s_j), PM2.5(s_i))$
across four equal-width longitude strips $[-180^{\circ}, -90^{\circ})$, $[-90^{\circ}, 0^{\circ})$, $[0^{\circ}, 90^{\circ})$, $[90^{\circ}, 180^{\circ}]$.  The symmetry is evident. 

\begin{figure}[ht]
  \centering
  \begin{subfigure}[b]{0.244\textwidth}
    \caption{Lon:$[-180^{\circ}, -90^{\circ})$}
    \includegraphics[width=1.1\textwidth]{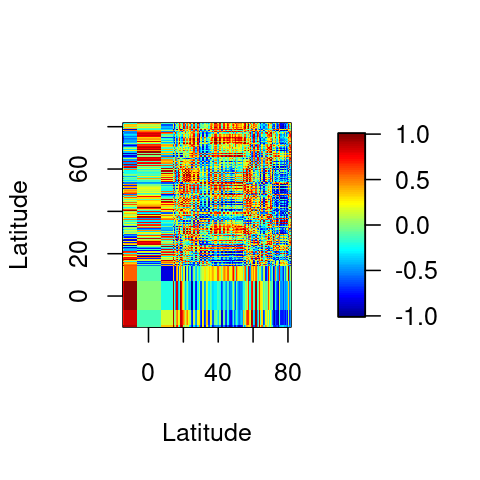}
  \end{subfigure}
  \begin{subfigure}[b]{0.24\textwidth}
    \caption{Lon:$[-90^{\circ}, 0^{\circ})$}
    \includegraphics[width=1.1\textwidth]{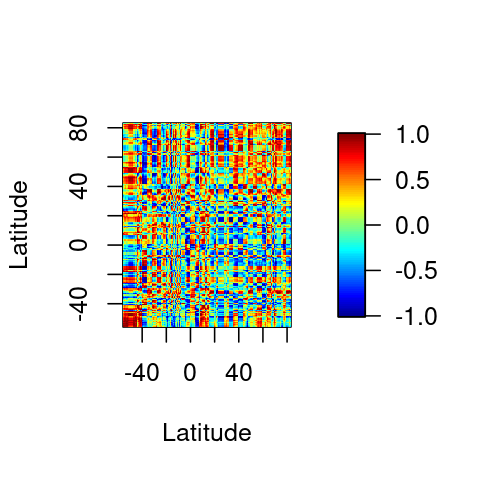}
  \end{subfigure}
  \begin{subfigure}[b]{0.24\textwidth}
    \caption{Lon:$[0^{\circ}, 90^{\circ})$}
    \includegraphics[width=1.1\textwidth]{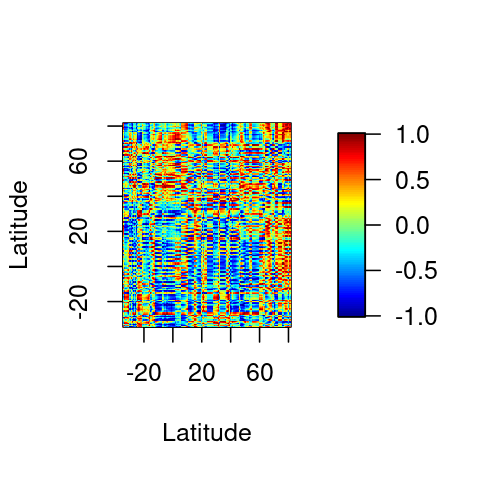}
  \end{subfigure}
  \begin{subfigure}[b]{0.24\textwidth}
    \caption{Lon:$[90^{\circ}, 180^{\circ}]$}
    \includegraphics[width=1.1\textwidth]{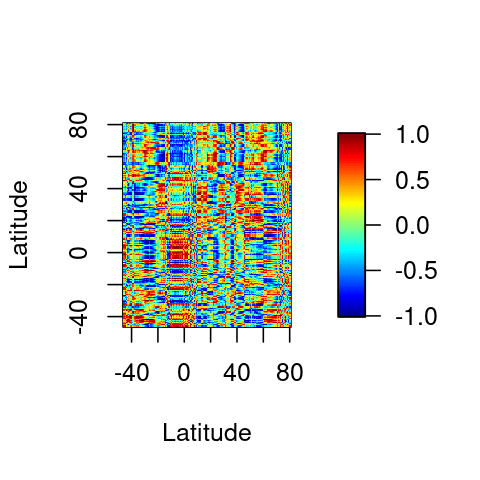}
  \end{subfigure}
  \caption{Empirical cross-correlation matrix plots for PM2.5 and Sea Salt across four longitude strips $[-180^{\circ}, -90^{\circ})$, $[-90^{\circ}, 0^{\circ})$, $[0^{\circ}, 90^{\circ})$, $[90^{\circ}, 180^{\circ}]$. All asymmetric.}
  \label{fig:cross_corr_SSvsPM25}
\end{figure}

Figure \ref{fig:cross_corr_SSvsPM25} is the empirical cross-correlation plots, displaying $corr(PM2.5(s_i), SS(s_j)) \neq corr(PM2.5(s_j), SS(s_i)) $ across four equal-width longitude strips. The asymmetry is prominent. 

In the following, 
Section \ref{sec:str_joint_Cov} looks into the structure of the joint covariance matrix where both the auto-covariance and cross-covariance reside. 
Section \ref{sec:properties} presents the properties of both the auto-correlation and cross-correlation. 
Section \ref{sec:criterion_model_Q} reviews mainstream multivariate models and analyses their capabilities to accommodate the asymmetric cross-covariance. 
Section \ref{sec:example} illustrates with a 1D simulated example.
Section \ref{sec:discuss} concludes the paper with a discussion.

\section{Structure of Joint Covariance Matrix}
\label{sec:str_joint_Cov}
Consider the data collected from the multivariate spatial phenomena described in Section \ref{sec:intro} arising from a multivariate discretely indexed spatial stochastic process $\{ (Y_1(s_i), \ldots, Y_p(s_i)): i = 1, \ldots, n \}$.

Let $\bmY = [\bmY^{T}_1(\cdot), \ldots, \bmY^{T}_l(\cdot), \ldots, \bmY^{T}_p(\cdot)]^{T} = [\bmY^{T}_1, \ldots, \bmY^{T}_i, \ldots, \bmY^{T}_n]^{T}$ represent a vector of $np$ random variables whose joint probability distribution $pr(\bmY)$ is supported on a product space $\Omega = \Omega_1 \times \ldots \times \Omega_{np}$, where each $\Omega_{*}$ is the support for one random variable $Y_l(s_i)$, $l = 1, \ldots, p$, and $i = 1, \ldots, n$.
The ``$\cdot$'' denotes all $n$ locations in the spatial domain $\mcalD$.

Here, $\bmY_l(\cdot)= [Y_l(s_1), \ldots, Y_l(s_n)]^T  \in \mbbR^n $ 
spanning across all $n$ locations for a particular component $l$, 
while $\bmY_i  = [Y_1(s_i), \ldots, Y_p(s_i)]^T \in \mbbR^p$ 
collecting all $p$ components at a particular location $s_i$. 
The equivalence of the joint distribution $pr(\bmY_1, \ldots, \bmY_{l}, \ldots, \bmY_p)$, where $\bmY_{l} \in \mbbR^n$, and $pr(\bmY_1, \ldots, \bmY_i, \ldots, \bmY_n)$, where $\bmY_i \in \mbbR^p$, 
is due to permutation invariant
property of a valid probability measure, see 
the FUNDAMENTAL THEOREM in 
\citet[p. 41]{kolmogorov2018foundations}.

Under the Gaussian process assumption, the second central moment, that is, the joint covariance matrix is the main modeling interest. 

Unlike univariate spatial stochastic processes (e.g., a single pollutant), whose joint covariance matrix encodes the spatial correlation of a single component spanning across all $n$ locations (and is therefore of dimension $n \times n$), the joint covariance matrix of a multivariate spatial stochastic process not only having blocks encoding the spatial correlation of a single component but also blocks encoding correlations of pairs of $p$ components across the spatial domain. 
Therefore, the joint covariance matrix is of dimension $np \times np$ and is denoted as $\HSigma_{np \times np}$.

Formally, 
the joint covariance matrix $\HSigma_{np \times np} = [cov(Y_l(s_i), Y_k(s_j))] \triangleq [C^{lk}(s_i, s_j)] $, where $C^{lk}(s_i, s_j)$ is a cross-covariance if $l \neq k$ and $i \neq j$. When $l = k$ but $i \neq j$, $C^{lk}(s_i, s_j)$ is a same-component auto-covariance, and when $l \neq k$, but $i = j$, $C^{lk}(s_i, s_j)$ is a same-location auto-covariance. $l, k = 1, \ldots, p$, $i, j = 1, \ldots, n$.


The joint covariance matrix for the random vector $[Y_1(s_1), \ldots, Y_p(s_1), \ldots, Y_1(s_n), \ldots, Y_p(s_n)]^T \in \mbbR^{np}$, consisting of the multivariate spatial stochastic process, is  
\begin{align}
\label{1st}
        \HSigma_{np \times np} = 
        \begin{bmatrix}
            [cov(\bmY(s_1), \bmY(s_1))]_{p \times p} & [cov(\bmY(s_1), \bmY(s_2))]_{p \times p} & \cdots & [cov(\bmY(s_1), \bmY(s_n))]_{p \times p}\\
            [cov(\bmY(s_2), \bmY(s_1))]_{p \times p} & [cov(\bmY(s_2), \bmY(s_2))]_{p \times p} & \cdots & [cov(\bmY(s_2), \bmY(s_n))]_{p \times p}\\
            \vdots & \vdots & \ddots & \vdots\\
            [cov(\bmY(s_n), \bmY(s_1))]_{p \times p} & [cov(\bmY(s_n), \bmY(s_2))]_{p \times p} & \cdots & [cov(\bmY(s_n), \bmY(s_n))]_{p \times p}
        \end{bmatrix}_{np\times np},
    \end{align}
where $\bmY(s_1) = [Y_1(s_1), Y_2(s_1), \ldots, Y_p(s_1)]^T \in \mbbR^p$.
    
Alternatively, after permutation, the joint covariance matrix of the random vector $[Y_1(s_1), \ldots, Y_1(s_n), \ldots, Y_p(s_1), \ldots, Y_p(s_n)]^T \in \mbbR^{np}$ is 
\begin{align}
\label{2nd}
        \HSigma_{np \times np} = 
        \begin{bmatrix}
            [cov(\bmY_1(\bms), \bmY_1(\bms))]_{n \times n} & [cov(\bmY_1(\bms), \bmY_2(\bms))]_{n \times n} & \cdots & [cov(\bmY_1(\bms), \bmY_p(\bms))]_{n \times n}\\
            [cov(\bmY_2(\bms), \bmY_1(\bms))]_{n \times n} & [cov(\bmY_2(\bms), \bmY_2(\bms))]_{n \times n} & \cdots & [cov(\bmY_2(\bms), \bmY_p(\bms))]_{n \times n}\\
            \vdots & \vdots & \ddots & \vdots\\
            [cov(\bmY_p(\bms), \bmY_1(\bms))]_{n \times n} & [cov(\bmY_p(\bms), \bmY_2(\bms))]_{n \times n} & \cdots & [cov(\bmY_p(\bms), \bmY_p(\bms))]_{n \times n}
        \end{bmatrix}_{np\times np},
    \end{align}
where $\bms = [s_1, \ldots, s_n]^T$, and $\bmY_1(\bms) = [Y_1(s_1), Y_1(s_2), \ldots, Y_1(s_n)]^T \in \mbbR^n$.

For a clear illustration, we expand one of the off-diagonal blocks $[cov(\bmY(s_1), \bmY(s_2))]_{p \times p}$ in Eq.~\eqref{1st} and $[cov(\bmY_1(\bms), \bmY_2(\bms))]_{n \times n}$ in Eq.~\eqref{2nd} to expose the \textit{asymmetry}. 

\begin{align}
\label{cross1}
        [cov(\bmY(s_1), \bmY(s_2))]_{p \times p} = 
        \begin{bmatrix}
            C^{11}(s_1, s_2) & C^{12}(s_1, s_2) & \cdots & C^{1p}(s_1, s_2)\\
            C^{21}(s_1, s_2) & C^{22}(s_1, s_2) & \cdots & C^{2p}(s_1, s_2)\\
            \vdots & \vdots & \ddots & \vdots\\
            C^{p1}(s_1, s_2) & C^{p2}(s_1, s_2) & \cdots & C^{pp}(s_1, s_2)
        \end{bmatrix}_{p\times p}, 
    \end{align}
where, in general, $cov(Y_1(s_1), Y_p(s_2)) \triangleq C^{1p}(s_1, s_2) \neq C^{p1}(s_1, s_2) \triangleq cov(Y_p(s_1), Y_1(s_2)) $.

\begin{align}
\label{cross2}
        [cov(\bmY_1(\bms), \bmY_2(\bms))]_{n \times n} = 
        \begin{bmatrix}
            C^{12}(s_1, s_1) & C^{12}(s_1, s_2) & \cdots & C^{12}(s_1, s_n)\\
            C^{12}(s_2, s_1) & C^{12}(s_2, s_2) & \cdots & C^{12}(s_2, s_n)\\
            \vdots & \vdots & \ddots & \vdots\\
            C^{12}(s_n, s_1) & C^{12}(s_n, s_2) & \cdots & C^{12}(s_n, s_n)
        \end{bmatrix}_{n\times n}, 
    \end{align}
where $cov(Y_1(s_1), Y_2(s_n)) \triangleq C^{12}(s_1, s_n) \neq C^{12}(s_n, s_1) \triangleq cov(Y_1(s_n), Y_2(s_1))$, in general.

Therefore, the off-diagonal blocks in Eq.~\eqref{1st} and \eqref{2nd} are generally asymmetric.

In contrast, all the blocks on the main diagonal are symmetric. We expand $[cov(\bmY(s_1), \bmY(s_1))]_{p \times p}$ in Eq.~\eqref{1st} and $[cov(\bmY_1(\bms), \bmY_1(\bms))]_{n \times n}$ in Eq.~\eqref{2nd} to expose the \textit{symmetry}. 

For the same-location auto-covariance matrix block,
\begin{align}
\label{same_loc}
        [cov(\bmY(s_1), \bmY(s_1))]_{p \times p} = 
        \begin{bmatrix}
            C^{11}(s_1, s_1) & C^{12}(s_1, s_1) & \cdots & C^{1p}(s_1, s_1)\\
            C^{21}(s_1, s_1) & C^{22}(s_1, s_1) & \cdots & C^{2p}(s_1, s_1)\\
            \vdots & \vdots & \ddots & \vdots\\
            C^{p1}(s_1, s_1) & C^{p2}(s_1, s_1) & \cdots & C^{pp}(s_1, s_1)
        \end{bmatrix}_{p\times p}, 
    \end{align}
where $cov(Y_1(s_1), Y_p(s_1)) \triangleq C^{1p}(s_1, s_1) \equiv C^{p1}(s_1, s_1) \triangleq cov(Y_p(s_1), Y_1(s_1)) $, by definition of covariance.

For the same-component auto-covariance matrix block,
\begin{align}
\label{same_var}
        [cov(\bmY_1(\bms), \bmY_1(\bms))]_{n \times n} = 
        \begin{bmatrix}
            C^{11}(s_1, s_1) & C^{11}(s_1, s_2) & \cdots & C^{11}(s_1, s_n)\\
            C^{11}(s_2, s_1) & C^{11}(s_2, s_2) & \cdots & C^{11}(s_2, s_n)\\
            \vdots & \vdots & \ddots & \vdots\\
            C^{11}(s_n, s_1) & C^{11}(s_n, s_2) & \cdots & C^{11}(s_n, s_n)
        \end{bmatrix}_{n\times n}, 
    \end{align}
where $cov(Y_1(s_1), Y_1(s_n)) \triangleq C^{11}(s_1, s_n) \equiv C^{11}(s_n, s_1) \triangleq cov(Y_1(s_n), Y_1(s_1))$, by definition of covariance.

Therefore, auto-covariance blocks, both the same-component and same-location, residing on the main diagonal of the joint covariance matrix $\HSigma_{np \times np}$, are always symmetric. 

\begin{figure}[htpb]
    \centering
    \includegraphics[width=0.45\linewidth]{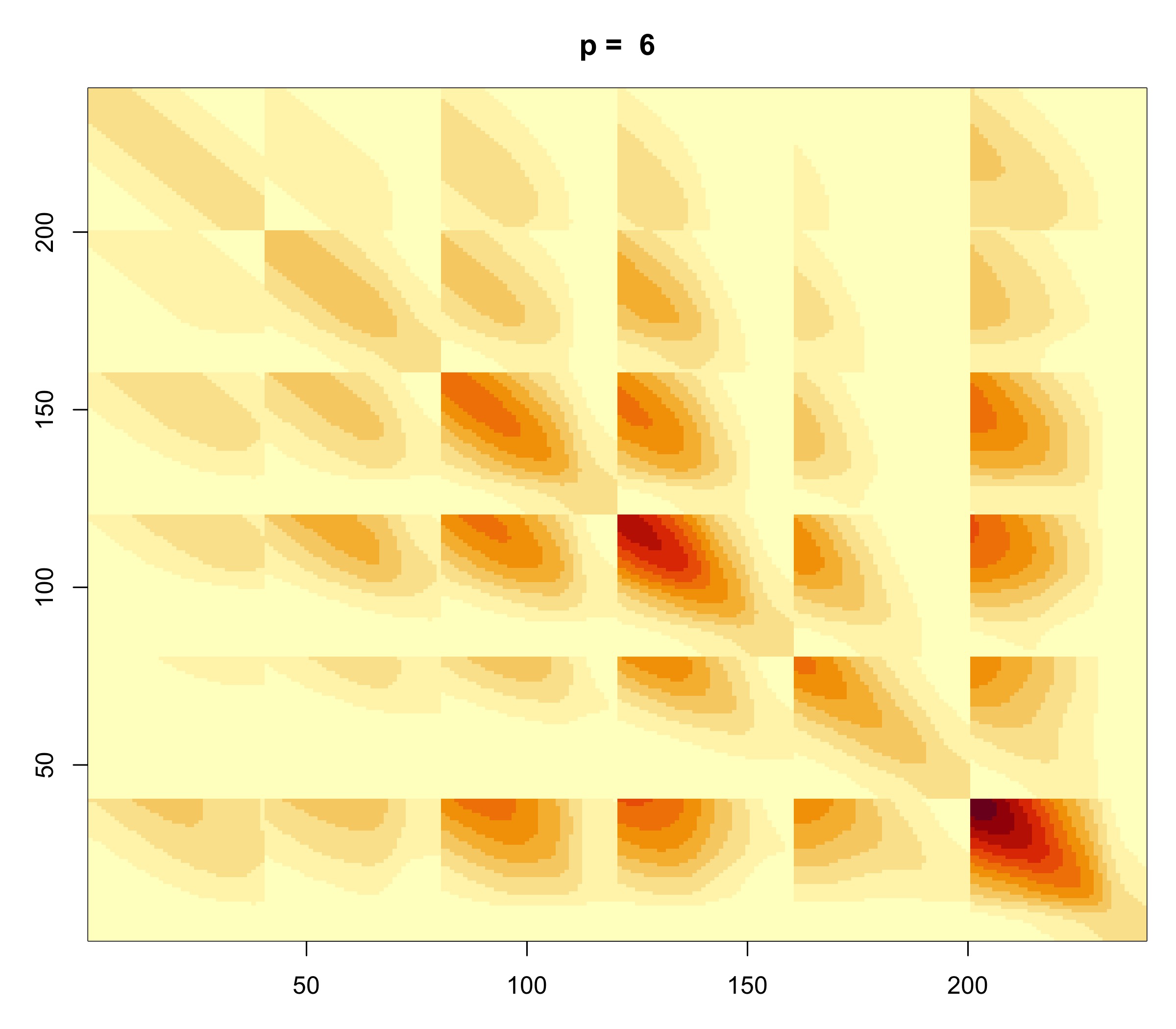}
    \caption{Simulated joint covariance matrix $\HSigma_{6*40 \times 6*40}$ for six components spanning 40 1D locations. All main diagonal blocks are symmetric, while each off-diagonal block is asymmetric. $\HSigma_{6*40 \times 6*40}$ must be positive definite and symmetric, therefore, the asymmetric off-diagonal blocks are symmetric across the main diagonal blocks.}
    \label{fig:simu_6}
\end{figure}

Figure \ref{fig:simu_6} is a simulated joint covariance matrix $\HSigma_{6*40 \times 6*40}$ for $p = 6$ (e.g., six types of pollutants) spanning a 1D spatial domain with $n = 40$ locations, where $s_i \in \mbbR^1$.

All the blocks on the main diagonal representing auto-covariance (organized as same-component auto here) are symmetric, while blocks on the off-diagonal representing cross-covariance are asymmetric.

Although off-diagonal blocks representing cross-covariance are asymmetric in general, the joint covariance matrix $\HSigma_{np \times np}$ itself must be positive definite and, therefore, symmetric as a whole. 
That is, the block $[cov(\bmY(s_1), \bmY(s_n))]_{p \times p}$ and $[cov(\bmY(s_n), \bmY(s_1))]_{p \times p}$ in Eq.~\eqref{1st}, as well as block $[cov(\bmY_1(\bms), \bmY_p(\bms))]_{n \times n}$ and $[cov(\bmY_p(\bms), \bmY_1(\bms))]_{n \times n}$ in Eq.~\eqref{2nd}, must be symmetric. 

The \textit{asymmetry} in the off-diagonal blocks must be \textit{symmetric} in the  $\HSigma_{np \times np}$.

\section{Properties of Auto-correlation and Cross-correlation}
\label{sec:properties}
The auto-/cross-correlation $Corr^{lk}(s_i, s_j) = C^{lk}(s_i, s_j) / \sqrt{C^{ll}(s_i, s_i)} \sqrt{C^{kk}(s_j, s_j)}$.

\subsection{Auto-correlation}
\begin{itemize}
    \item Symmetric, see Eq.~\eqref{same_loc}, \eqref{same_var}, and Fig. \ref{fig:auto_corr_PM25}.
    \item The main diagonal values $Corr^{ll}(s_i, s_i)$ of the auto-correlation matrix have the largest magnitude of 1, where $l= 1, \ldots p$, and $i = 1, \ldots, n$, see Fig. \ref{fig:auto_corr_PM25}.
    \item The magnitude of the off-diagonal values in the auto-correlation matrix must be less than or equal to the main diagonal value of 1, specifically,
    \begin{itemize}
        \item $|Corr^{lk}(s_i, s_i)| \leq |Corr^{ll}(s_i, s_i)|$, for same-location auto, $l \neq k$
        \item $|Corr^{ll}(s_i, s_j)| \leq |Corr^{ll}(s_i, s_i)|$, for same-component auto, $i \neq j$, see Fig. \ref{fig:auto_corr_PM25}.
    \end{itemize} 
    \item Auto-correlation can be both positive and negative, see Fig. \ref{fig:auto_corr_PM25}.
\end{itemize}

\subsection{Cross-correlation}
\begin{itemize}
    \item Asymmetric in general, see Eq.~\eqref{cross1}, \eqref{cross2}, and Fig. \ref{fig:cross_corr_SSvsPM25}.
    \item The largest magnitude (i.e., 1) of cross-correlation does \textit{not} necessarily occur on the main diagonal of the cross-correlation matrix and can occur in the off-diagonal areas, see Fig. \ref{fig:cross_corr_SSvsPM25}.
    \item Magnitudes of the off-diagonal values are \textit{not} necessarily less than the main diagonal values of the cross-correlation matrix. Specifically, the inequality $|Corr^{lk}(s_i, s_j)| \leq |Corr^{ll}(s_i, s_j)|$ (for Eq.\eqref{cross1}) or $|Corr^{lk}(s_i, s_j)| \leq |Corr^{lk}(s_i, s_i)|$ (for Eq.\eqref{cross2}),
    does \textit{not} hold in general, see Fig \ref{fig:cross_corr_SSvsPM25}.
    \item The magnitude $|Corr^{lk}(s_i, s_j)|$ of a cross-correlation is bounded by the product of the magnitudes of the main diagonal values of two auto-correlations, i.e., $|Corr^{lk}(s_i, s_j)| \leq |Corr^{ll}(s_i, s_i))| |Corr^{kk}(s_j, s_j))| \equiv 1 $ \citep[p.~150]{wackernagel2013multivariate}.
    \item Cross-correlation can be both positive and negative, see Fig. \ref{fig:cross_corr_SSvsPM25}.
\end{itemize}

\section{A Criterion for Assessing Model Quality}
\label{sec:criterion_model_Q}
From the structure of the joint covariance matrix $\HSigma_{np \times np}$ in Eq.~\eqref{1st}, \eqref{2nd} and Fig.~\ref{fig:simu_6}, we see that the off-diagonal blocks representing the cross-covariance matrix are asymmetric in general. 

Therefore, the capability to restore the asymmetry in the off-diagonal cross-covariance matrix blocks of $\HSigma_{np \times np}$ serves as one criterion for assessing the quality of the multivariate spatial models.

Below, we review several mainstream modeling methods for constructing the joint covariance matrix $\HSigma_{np \times np}$ and analyze their capability of incorporating the asymmetric cross-covariance. 

\subsection{Intrinsic Correlation Model}
\label{sec:intrin}
The simplest construction method is the \textit{intrinsic correlation model} \citep[p.~155]{wackernagel2013multivariate} or \textit{separable model} \citep[p.~263]{banerjee2014hierarchical}, which arises from a linear model of coregionalization framework, see \citet[p.~172-173]{wackernagel2013multivariate}.

The idea is decomposing a covariance matrix block $[cov(\bmY(s_i), \bmY(s_j))]_{p \times p}$ at a given pair of locations, denoted as $\bm{C}(s_i, s_j)$, into a product of two parts: one is a correlation reflecting the pure spatial correlation between the pair of locations for a univariate spatial process, denoted as $\rho(s_i, s_j)$, and the other is a pure variance-covariance matrix among $p$ components, denoted as $\bm{V}_{p \times p}$, in which $V^{lk} = cov(Y_l, Y_k)$. 

Together, the covariance matrix block at the given pair of locations is represented as
    \begin{align}
    \label{LMC_C}
        \bm{C}(s_i, s_j) &= \rho(s_i, s_j)\bm{V}_{p \times p}.
    \end{align}
Therefore, the desired joint covariance matrix $\HSigma_{np \times np}$ is a Kronecker product of a spatial correlation matrix $\bm{H}_{n \times n}$ and a variance-covariance matrix $\bm{V}_{p \times p}$ among $p$ components, that is,
    \begin{align}
    \label{kroneck}
        \HSigma_{np \times np} =  \bm{H}_{n \times n} \otimes \bm{V}_{p \times p},
    \end{align}
where each $\bm{H}_{i, j}$ corresponds to the spatial correlation $\rho(s_i, s_j)$, that is, $\bm{H}_{ij} = \rho(s_i, s_j)$. 

Here, $V^{lk} = cov(Y_l, Y_k) \equiv cov(Y_k, Y_l) = V^{kl}$, by covariance definition, hence the matrix $\bm{V}_{p\times p}$ is symmetric. 
Then, each off-diagonal block $[\bm{C}(s_i, s_j)]_{p \times p}$ in $\HSigma_{np \times np}$ is a spatial correlation scalar $\rho(s_i, s_j)$ times a symmetric matrix $\bm{V}_{p\times p}$, making the off-diagonal block $[\bm{C}(s_i, s_j)]_{p \times p}$ symmetric as a whole. 

To illustrate clearly, we substitute the generic Eq.~\eqref{cross1} with 
Eq.~\eqref{LMC_C}, 
\begin{align}
        [cov(\bmY(s_1), \bmY(s_2))]_{p \times p} &= 
        \begin{bmatrix}
            C^{11}(s_1, s_2) & C^{12}(s_1, s_2) & \cdots & C^{1p}(s_1, s_2)\\
            C^{21}(s_1, s_2) & C^{22}(s_1, s_2) & \cdots & C^{2p}(s_1, s_2)\\
            \vdots & \vdots & \ddots & \vdots  \nonumber \\
            C^{p1}(s_1, s_2) & C^{p2}(s_1, s_2) & \cdots & C^{pp}(s_1, s_2)
        \end{bmatrix}_{p\times p} \\
        &= \rho(s_1, s_2) \begin{bmatrix}
        \label{eq:symm_lmc}
            V^{11} & V^{12} & \cdots & V^{1p}\\
            V^{21} & V^{22} & \cdots &  V^{2p}\\
            \vdots & \vdots & \ddots & \vdots\\
            V^{p1} &  V^{p2} & \cdots & V^{pp}
        \end{bmatrix}_{p\times p}.
    \end{align}
Here, $ V^{1p} \equiv V^{p1} $.

Therefore, this model lacks a mechanism to accommodate the asymmetric cross-covariance in the off-diagonal blocks of $\HSigma_{np \times np}$.

\subsection{Kernel Convolution Approach}
\label{sec:kenel}
\citet{ver1998constructing} proposed a moving average approach for generating the desired joint covariance matrix $\HSigma_{np \times np}$. This method is also known as \textit{kernel convolution}. 

The idea is to generate each component field of the multivariate random field individually from a common underlying hidden process.

Let $g(\cdot)$ be a standard Gaussian process $\mathcal{N}(0, 1)$ whose correlation between pairs of locations is $\rho(s_i, s_j)$. 
$k_l(\cdot)$ is a square integrable kernel function on $\mbbR^2$, where $l = 1, 2, \ldots, p$. 
Define
\begin{align}
\label{eq:Y_l_kernel}
    Y_l(s_i) = \sigma_l \int k_l(s_i - t) g(t) dt, \quad l = 1, 2, \ldots, p; \quad
    i = 1, \ldots, n.
\end{align}
then the joint random vector $\bmY \in \mbbR^{np}$ is still a mean zero Gaussian process. 

The $(lr)^{th}$ cross-covariance block $ [C^{lr}(\bms, \bms)]_{n \times n}$, where $\bms = [s_1, \ldots, s_n]^T$, of the joint covariance matrix $\HSigma_{np \times np} $ having the $(i, j)^{th}$ element as  
\begin{align}
\label{eq:C_lr_kernel}
    C^{lr}(s_i, s_j) = \sigma_l \sigma_r \int \int k_l(s_i - t) k_r (s_j - t) \rho(t - t') d t d t'.
\end{align}

Here, the kernel function $k_l(\cdot)$ and $k_r(\cdot)$ can accommodate the asymmetry. Specifically, by introducing a shift parameter $\Delta$ to the location separation lag $s-t \triangleq h$. 
Unlike Euclidean distance $\|h\|$ having no direction, separation lag $h$ has directions,
consequently, $h_{ij} - \Delta \neq h_{ji} - \Delta$. 
Therefore, the $(i, j)^{th}$ element $C^{lr}(s_i, s_j)$ and $(j, i)^{th}$ element $C^{lr}(s_j, s_i)$ of the $(lr)^{th}$ 
cross-covariance matrix block $ [C^{lr}(\bms, \bms)]_{n \times n}$ is not equal, hence the asymmetry.

The idea of inducing the asymmetry through shifting was proposed by \citet{li2011approach}.


\subsection{Multivariate Mat\'ern Approach}
\label{sec:multimatern}
\citet{gneiting2010matern} introduced a multivariate \matern{} method to construct the $\HSigma_{np \times np}$ for the multivariate spatial stochastic processes or multivariate random field. 

Specifically, the auto-correlation for each univariate process of the multivariate random field is modeled using \matern{} correlation; meanwhile, the cross-correlation for a given pair of processes is also modeled using \matern{}. That is, 
\begin{align*}
    Corr^{ll}(h) &= M(h; \nu_l, \kappa_l) \\
    Corr^{lk}(h) &= \beta_{lk} M(h; \nu_{lk}, \kappa_{lk}),
\end{align*}
where $M(\cdot)$ represents the \matern{} correlation, with parameter $\nu$ controlling the small-scale smoothness near the origin, and $\kappa$ controlling the rate of the correlation decay at a large spatial scale \citep[p.~32]{stein1999interpolation}. Separation lag $h \in \mbbR^d$.

Here, $\beta_{lk}$ is the cross-correlation between two components $l,k$ regardless of location.
Since \matern{} correlation is a kernel, the corresponding kernel matrix must be positive definite (and thus symmetric). Therefore, the cross-correlation block matrix is a scalar $\beta_{lk}$ times a symmetric matrix, which makes the block matrix symmetric. 

However, by inducing a shift parameter $\Delta$ to the spatial separation lag, the $(i, j)^{th}$ element of the kernel matrix is not equal to its $(j, i)^{th}$ element due to $h_{ij} - \Delta \neq h_{ji} - \Delta$. Consequently, the cross-correlation matrix block becomes asymmetric.

Therefore, this model has the mechanism to accommodate the asymmetry in the cross-covariance matrix blocks.

\subsection{Conditional Modeling Approach}
\label{sec:condi}
The main idea of this class of model is to construct the desired joint covariance matrix $\HSigma_{np \times np}$ or the joint precision matrix $\HSigma_{np \times np}^{-1}$ through the specification of conditional mean and conditional variance.  
The idea of obtaining the joint covariance from conditional mean and covariance dates back as early as in \citet{yule1907theory}, and is also in \citet[p. 370-372]{bishop2007pattern}.

\subsubsection{Mardia's Conditional} 
\label{sec:mardi}
\citet{mardia1988multi} proposed a conditional approach to model the joint precision matrix $\HSigma_{np \times np}^{-1}$. 

Under Gaussian data assumption, let $\bm{Y}_i = [Y_1(s_i), Y_2(s_i), \ldots, Y_p(s_i)]^T \in \mbbR^p$. 
By modeling each $\mbfY_i$ conditionally as
\begin{small}
\begin{align}
\label{mardiaeq}
    \mbbE[\mbfY_i | \mbfY_{-i}] &= \bfmu_i + \sum_{j \in \mcalN(i)} \bfbeta_{ij} (\mbfY_i - \bfmu_j), \quad
    Var[\mbfY_i | \mbfY_{-i}] = \mathbf{\Gamma}_i,
\end{align}
\end{small}
the joint precision matrix $\HSigma_{np \times np}^{-1}$ = $\{ \mbox{block diag}(\mathbf{\Gamma}_i)^{-1} \} \{ \mbox{block} (- \bfbeta_{ij}) \}$,
provided the symmetric condition (i.e., $\bm{\Gamma}_i ^{-1} \bfbeta_{ij} = \bfbeta_{ji}^T \bm{\Gamma}_j ^{-1}$) and the positive definite condition (i.e.,  
$[\mbox{block}(- \bfbeta_{ij})]$ being positive definite) are satisfied. 

Since the method models the joint precision matrix $\HSigma_{np \times np}^{-1}$ rather than the joint covariance matrix $\HSigma_{np \times np}$ where the asymmetric cross-covariance matrix blocks reside, the asymmetry of the cross-covariance can not be readily accommodated.

\subsubsection{Cressie's Conditional}
Unlike \citet{mardia1988multi} which constructs joint precision matrix $\HSigma_{np \times np}^{-1}$,
\citet{cressie2016multivariate} constructs joint covariance matrix $\HSigma_{np \times np}$.

Assuming Gaussian data, denote $Y_q(s)$ as the component $q$ at location $s$.
By modeling the conditional mean of $Y_q$ and the conditional covariance of $Y_q$ at two locations,
\begin{small}
\begin{align}
\label{eq:col-wise_multi}
    \mbbE(Y_q(s_i) | \{\bmY_r(\cdot): r = 1, 2, \cdots, (q-1)\}) &= \sum_{r = 1} ^{(q-1)} \int_D b_{qr} (s_i, v) Y_r(v) dv; \nonumber \\
    cov(Y_q(s_i), Y_q(s_j) | \{\bmY_r(\cdot): r = 1, 2, \cdots, (q-1) \}) &= C_{q | (r < q)}(s_i, s_j); \quad s_i, s_j \in \mcalD ,    \nonumber
\end{align}  
\end{small}
one obtains $\HSigma_{np \times np}$.

Under a bivariate situation ($p = 2$), the $\HSigma_{np \times np}$ has $\Sigma_{11}$ (of dimension $n \times n$) as the leading diagonal block, and $\bm{B} \Sigma_{11}$ as the off-diagonal blocks, where matrix $\bm{B}$ is the $b_{qr} (s_i, v)$ function evaluated on all pair of spatial locations in $\mcalD$, see \citet[p.~160-161]{cressie2011statistics}. 

Similar to the method in Section \ref{sec:kenel}, the $b_{qr} (s_i, v)$ function can be used to accommodate the asymmetric cross-covariance matrix by introducing a shift parameter $\Delta$. 


\subsection{Remark}
The mainstream multivariate models can be broadly categorized into two types. One is the unconditional type, which directly models the off-diagonal cross-covariance matrix. 
Examples include the \textit{intrinsic correlation model} (Section \ref{sec:intrin}), the \textit{kernel convolution model} (Section \ref{sec:kenel}), and the \textit{multivariate Mat\'ern model} (Section \ref{sec:multimatern}). 
The other type consists of conditional modeling methods discussed in Section \ref{sec:condi}.

For the unconditional type, the direct models can be further divided into two kinds, 
one is modeling the off-diagonal cross-covariance matrix in the form of Eq.~\eqref{cross1}, which is of dimension $p \times p$; examples include the \textit{intrinsic correlation model}.
The other is modeling the off-diagonal cross-covariance matrix in the form of Eq.~\eqref{cross2}, which has a dimension of $n \times n$; examples include the \textit{kernel convolution}, and \textit{multivariate Mat\'ern}. 

The first kind of $p \times p$ cross-covariance matrix expands its dimension by encoding correlation between pairs of components while keeping the spatial correlation fixed for a given pair of locations. Since $corr(Y_l, Y_k) \equiv corr(Y_k, Y_l)$, and currently no strategy to break this equivalence while maintaining scientific interpretability, this kind of direct construction lacks a mechanism to induce an asymmetry in the off-diagonal cross-covariance matrix block, see Eq.~\eqref{eq:symm_lmc}.

In contrast, 
the second kind of $n \times n$ cross-covariance matrix expands its dimension by encoding the correlation at different spatial locations for a given pair of components. 
By inducing a shifting parameter to the spatial separation leg, the $(i, j)^{th}$ element of the $n \times n$ cross-covariance matrix is not the same as the $(j, i)^{th}$ element, where $i, j = 1, \ldots, n$. 
Therefore, this second kind of direct construction method is relatively more effective at accommodating asymmetry.

\section{Example}
\label{sec:example}
This section presents a 1D simulation using the method in \citet{cressie2016multivariate}. 

The simulated spatial domain $\mcalD$ is $[-10, 10]$, discretized with a grid size of 0.1, resulting in 200 locations ($n = 200$). Each $s_i \in \mcalD \subset \mbbR^1$. 
The simulation models a tri-variate process ($p = 3$), such as three pollutants, each spanning the spatial domain $\mcalD$, hence three random fields. 

The goal is to predict the true process values at the first 50 locations of the first field, given the remaining noisy observations in the same field and 400 observations from the other two fields (200 per field).

The prediction results are compared between the model with a shift parameter $\Delta$, which accounts for the asymmetric cross-correlation, and the model without it.

\begin{figure}[ht]
    \centering
    \includegraphics[width=0.65\linewidth]{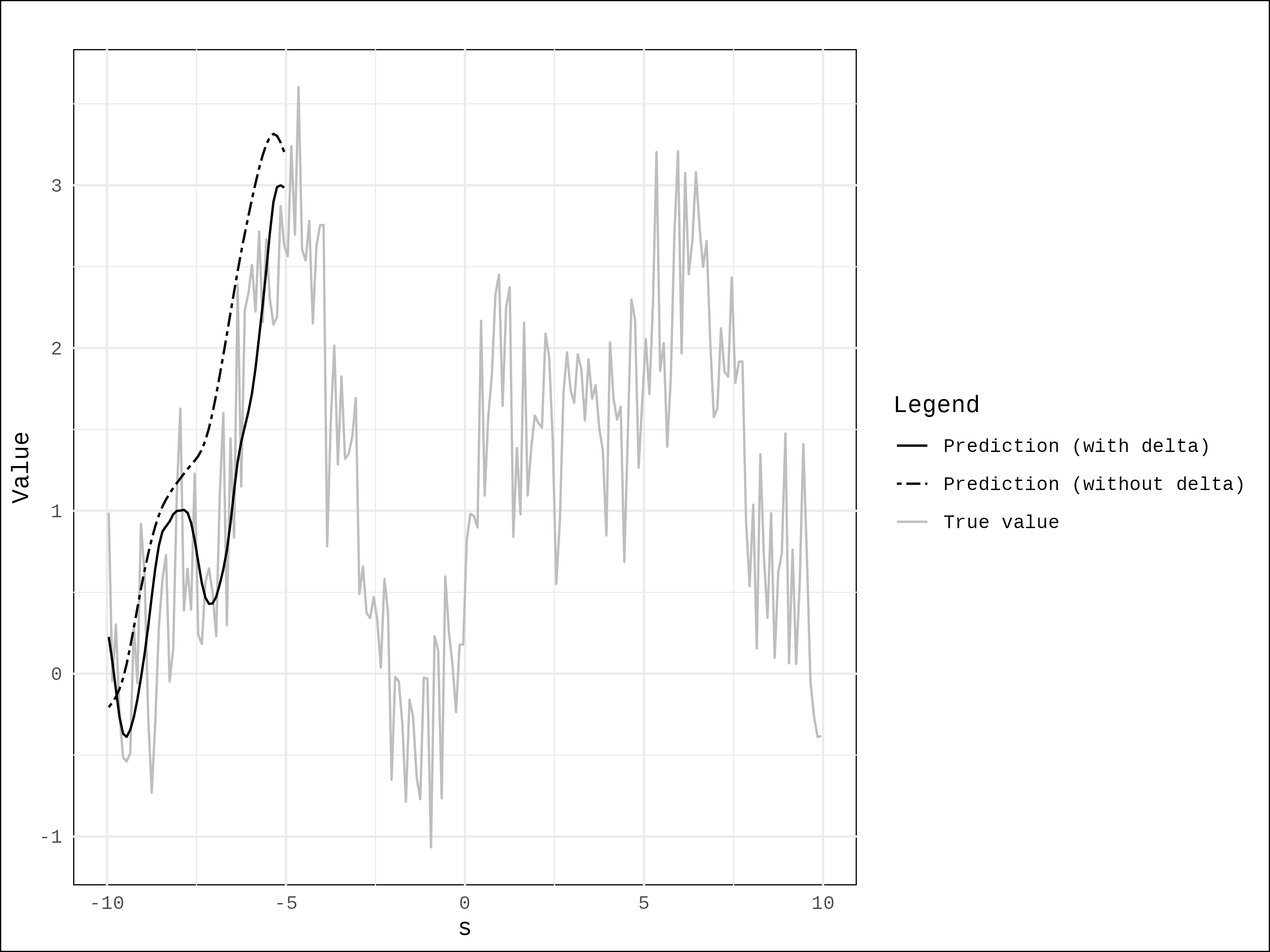}
    \caption{Prediction results for the first 50 locations of the first field, given the noisy observations from the remaining locations in the first field and the 400 observations from the other two fields.}
    \label{fig:co-krig}
\end{figure}

Figure \ref{fig:co-krig} shows the model with $\Delta$, which accounts for the asymmetry, produces more accurate prediction results (solid line) than the one without it (the dashed line). 

Table \ref{tab:compare} compares model accuracy using two metrics: Mean Absolute Error (MAE) and Root Mean Square Error (RMSE).
The model with the shift parameter $\Delta$ has lower MAE and RMSE than the one without.

\begin{table}[htbp]
    \centering
    \caption{Compare MAE and RMSE between the Model with $\Delta$ and the One without $\Delta$}
    \begin{tabular}{c cc}
         & MAE & RMSE \\ \hline 
      with $\Delta$  & 0.245 &  0.289 \\
    without $\Delta$ & 0.706 & 0.740
    \end{tabular}
    \label{tab:compare}
\end{table}

\section{Discussion}
\label{sec:discuss}
Multivariate spatial phenomena are ubiquitous, spanning domains such as climate, pandemics, air quality, and the social economy.
Quantities of interest not only interact with themselves at nearby regions but also interact with other quantities at nearby locations, as reflected by cross-correlation (or cross-covariance). 

The cross-correlation is generally asymmetric, and the capability to capture this asymmetry is one of the criteria for assessing model quality.

Admittedly, asymmetric cross-covariance is just one of the features characterizing the multivariate spatial data class. 
Other factors, such as computational efficiency, are also important and serve as criteria as well for assessing the model's quality in terms of computational feasibility and performance.

Methods such as \textit{intrinsic correlation model} (Sec. \ref{sec:intrin}) and \textit{Mardia's conditional} (Sec. \ref{sec:mardi}) may not readily capture the asymmetric cross-covariance in the off-diagonal blocks of $\HSigma_{np \times np}$, but they do offer tangible computational efficiency. 

Concretely, due to the Kronecker product structure in Eq.~\eqref{kroneck}, the intrinsic correlation model 
significantly eases the computation burden for computing the determinant $|\HSigma_{np \times np}|$ and inverse $\HSigma_{np \times np}^{-1}$, since both operations reduce to the Kronecker product of two much smaller matrices, specifically, $|\HSigma_{np \times np}| = |\bm{H}|^p  \otimes |\bm{V}|^n$, and $\HSigma_{np \times np}^{-1} = \bm{H}^{-1} \otimes \bm{V}^{-1}$. 

For Mardia's conditional approach, since the conditional mean in Eq.~\eqref{mardiaeq} only regresses on the values within a neighborhood of $s_i$, the resulting precision matrix  $\HSigma_{np \times np}^{-1}$ naturally embodies structural sparsity, enhancing computational efficiency.

Asymmetric cross-covariance matrix blocks reside in the joint covariance matrix 
$\HSigma_{np \times np}$, while the sparsity is present in the joint precision matrix $\HSigma_{np \times np}^{-1}$. 
Methods that can simultaneously address asymmetric cross-covariance and maintain computational efficiency (e.g., a sparse precision matrix) are increasingly needed.

\section*{Acknowledgments}
The author was supported by the Alan Turing Institute doctoral studentship under EPSRC grant EP/N510129/1.

\bibliography{main} 

\end{document}